# Cataclysmic Variables and Other Compact Binaries in the Globular Cluster NGC 362: Candidates from *Chandra* and *HST*

B. Margon[*], B. Beck-Winchatz[†], L. Homer[**], D. Pooley[‡], C. G. Bassa[§], S. F. Anderson[**], W. H. G. Lewin[¶], F. Verbunt[‖], A. K. H. Kong[††], and R. M. Plotkin[§§]

[*]*Department of Astronomy & Astrophysics, University of California, Santa Cruz, CA 95064, USA*
[†]*Scientific Data Analysis and Visualization Program, DePaul University, Chicago, IL 60614, USA*
[**]*Department of Astronomy, University of Washington, Box 351580, Seattle, WA 98195, USA*
[‡]*Eureka Scientific, 5248 Valley View Road, El Sobrante, CA 94803, USA*
[§]*Jodrell Bank Centre for Astrophysics, School of Physics and Astronomy, University of Manchester, Manchester M13 9PL, UK*
[¶]*Kavli Institute for Astrophysics and Space Research, Massachusetts Institute of Technology, Cambridge, MA 02139, USA*
[‖]*Astronomical Institute, Utrecht University, PO Box 80000, 3508 TA Utrecht, The Netherlands*
[††]*Institute of Astronomy and Department of Physics, National Tsing Hua University, Hsinchu 30013, Taiwan*
[§§]*Astronomical Institute "Anton Pannekoek," University of Amsterdam, Science Park 904, 1098 XH, Amsterdam, The Netherlands*

**Abstract.** Highly sensitive and precise X-ray imaging from *Chandra*, combined with the superb spatial resolution of *HST* optical images, dramatically enhances our empirical understanding of compact binaries such as cataclysmic variables and low mass X-ray binaries, their progeny, and other stellar X-ray source populations deep into the cores of globular clusters. Our *Chandra* X-ray images of the globular cluster NGC 362 reveal 100 X-ray sources, the bulk of which are likely cluster members. Using *HST* color-magnitude and color-color diagrams, we quantitatively consider the optical content of the NGC 362 *Chandra* X-ray error circles, especially to assess and identify the compact binary population in this condensed-core globular cluster. Despite residual significant crowding in both X-rays and optical, we identify an excess population of Hα-emitting objects that is statistically associated with the *Chandra* X-ray sources. The X-ray and optical characteristics suggest that these are mainly cataclysmic variables, but we also identify a candidate quiescent low mass X-ray binary. A potentially interesting and largely unanticipated use of observations such as these may be to help constrain the macroscopic dynamic state of globular clusters.

## 1. INTRODUCTION

It has been recognized for decades that close binary stars play a key role in the dynamical evolution of globular star clusters, quite aside from the intrinsic interest in these systems individually. A dozen or two close binaries in a cluster can store as much binding energy as the sum of the orbital energy of all $10^5$ single cluster stars, and this energy can be extracted through encounters. Binaries are therefore storage batteries that can dominate the kinematic evolution of the cluster. Indeed, this stored energy may eventually inhibit the collapse of the cluster. Interestingly, the actual kinematic state of many clusters is not always obvious from the available instantaneous snapshot of feasible observations; thus even a simple census of close binaries may be one of the best empirical indicators of the macroscopic state of the cluster [1].

Although variations of this theoretical argument have been recognized for many years, empirical progress has been painfully slow until the launch of *HST* and *Chandra*. Indeed, a common refrain of the 1980s and early 1990s in this discipline was "where are the close binary stars in globular clusters?" This current conference happens to occur with a few days of the 30$^{th}$ anniversary of the appearance of a now amusingly naïve paper that discussed the probable discovery of the first optically-identified close binary in a globular cluster [2]. Even 15 years ago, a paper which boldly predicted ~$10^2$ close binaries in rich clusters such as 47 Tuc and ω Cen [3] was considered quite daring. Today, largely with the aid of spaceborne observatories, we know of well over $10^3$ cluster X-ray sources that are likely all close binary stars, residing in almost $10^2$ distinct clusters (Figure 1).

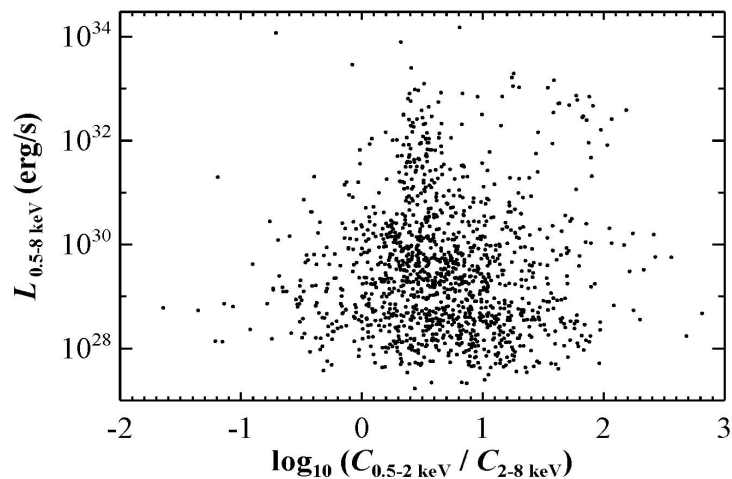

**FIGURE 1.** An X-ray "Hertzsprung-Russell" diagram (luminosity versus "color") for 1,500 X-ray sources detected by *Chandra* in the direction of 77 galactic globular clusters [1]. Virtually all are likely to be cluster members. Contrast this with the one such object known 30 years ago [2].

The identification of this large sample has permitted significant progress in understanding the formation mechanisms of close binaries in the exotic cluster environment. In particular, there is a very strong correlation of the number of X-ray sources in a given cluster with a measure of the cluster's stellar encounter rate [4] (Figure 2). In some sense, however, this relationship raises as many new questions as it answers. We now understand that the cluster X-ray population is complex and diverse, including at least four distinct types of objects, namely cataclysmic variables, active binaries (both chromospheric and magnetic), quiescent low-mass X-ray binaries, and millisecond pulsars. Therefore, why should the simple, elegant relationship of Figure 2 prevail amongst this heterogeneous mixture? Further, at least in the very dense clusters, binary formation and destruction terms must surely compete in complicated ways. Dense clusters are therefore particularly interesting for X-ray observations, but are also the most difficult place to obtain optical observations and thus characterize the X-ray emitting population.

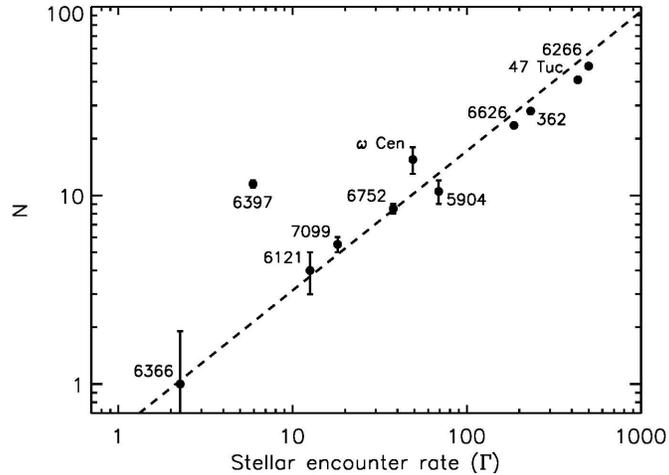

**FIGURE 2.** The relationship between the number of X-ray sources in selected globular clusters and the cluster stellar encounter rate, Γ [4]. The observations of NGC 362 described here are plotted near the upper right of the diagram.

## 2. OBSERVATIONS

The dense clusters 47 Tuc and ω Cen have been well-observed in X-rays, but if one looks for further cases there are unfortunately few others comparably favorable for detailed observation. Here we discuss *Chandra* and *HST* observations of NGC 362, another dense cluster, but unfortunately, at *(m-M)*~14.6, about twice the distance of the more famous siblings. As a result, the cluster core and nearby areas are extremely crowded. The X-ray data consist of 79 ks total exposure with the *Chandra* ACIS, and the corresponding *HST* optical data are ACS exposures over multiple epochs in three filters: $B_{425}$, $r_{625}$, and $H\alpha_{658}$. This paper is meant as a preliminary progress report on an ongoing analysis.

The X-ray results are shown in Figure 3, where we see 101 distinct point sources. The number of unrelated sources accidentally superposed on the clusters, estimated by several different methods, would appear to be very small, so virtually all these X-ray objects are cluster members. After adjusting for the appropriate X-ray flux threshold, the count of sources in our NGC 362 data fits beautifully onto the Pooley *et al*. relationship of Figure 2.

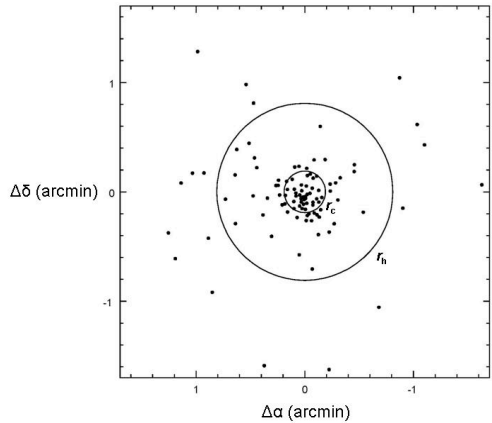

**FIGURE 3.** *Chandra* X-ray image of the center of the rich cluster NGC 362. The cluster core radius and half-mass radius are indicated. Only a handful of the 101 detected point sources are likely to be unrelated superpositions.

It is possible to begin characterization of the NGC 362 X-ray sources with X-ray data alone, using the X-ray "luminosity-color" relationship discussed by Pooley & Hut [5]. Our NGC 362 data in this system are shown in

Figure 4. While sources in Regions II and III of the diagram have been found to be of several different classes, Region I is known to effectively segregate quiescent low-mass X-ray binary (qLMXB) stars [5]. Indeed we see in Figure 4 one such bright X-ray source, located within a few arcsec of the cluster core, and we suggest this is likely a classical qLMXB.

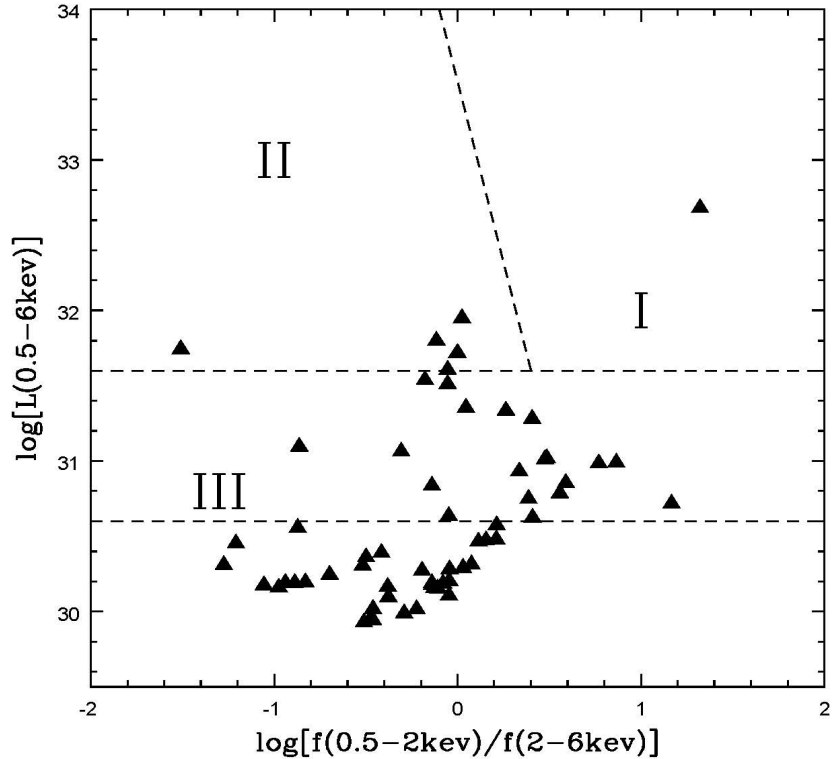

**FIGURE 4.** The brightest *Chandra* X-ray sources in NGC 362 plotted in an X-ray "luminosity-color" diagram. Sources located in Region I have been shown [5] to be qLMXBs with high probability.

X-ray data alone cannot effectively characterize the sources in Regions II and III of Figure 4, but our complementary HST optical data are of great value here. Given the tremendous spatial resolution of HST images, at the more modest Chandra resolution (of order 1" when both positional and astrometric uncertainties are included) there are numerous optical objects that are candidate counterparts to each X-ray source. We show one such example field in Figure 5, where the ACS data clearly detect and resolve several dozen stars in a typical X-ray error circle.

Although it is obvious optical identifications and source characterization will not be possible based solely on positional coincidence, the HST multicolor photometry adds yet another important dimension to the analysis. In the left panel of Figure 6 we display the photometry for more than 3,100 objects detected within 1" of a *Chandra* X-ray position in NGC 362. (Note that the vast majority of these objects must be unrelated to the X-ray sources, but merely superposed on the field). The oval draws attention to a subset of objects with blue colors and Hα excesses when compared to the typical object in the region. In the right panel, we show the identical data, but have also plotted with star symbols *Sloan Digital Sky Survey* [6] photometry, converted to the *HST* system via SYNPHOT, of ~$10^2$ spectroscopically-confirmed field cataclysmic variable stars [7]. The SDSS CVs are observed to be reasonably well segregated to the blue, Hα excess corner of the diagram. This is strong, although certainly not conclusive, evidence that many of the *HST* objects within the oval in the left panel are in fact CVs that are the counterparts of the *Chandra* X-ray sources. One of these objects, for example, is located in the field of X-ray Source 6 (Figure 5), within about an arcsec of the center of the X-ray positional error circle.

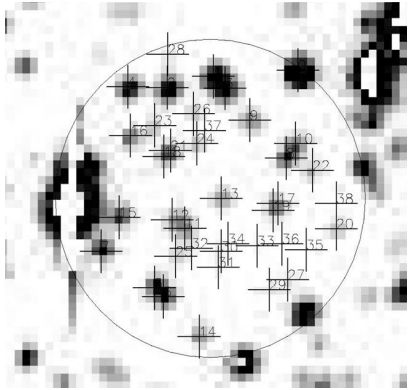

**FIGURE 5.** An HST B-band ACS exposure of the position of one *Chandra* X-ray source ("Source 6"); several dozen candidate optical counterparts are visible in the 1" radius error circle which combines X-ray positional and astrometric uncertainties.

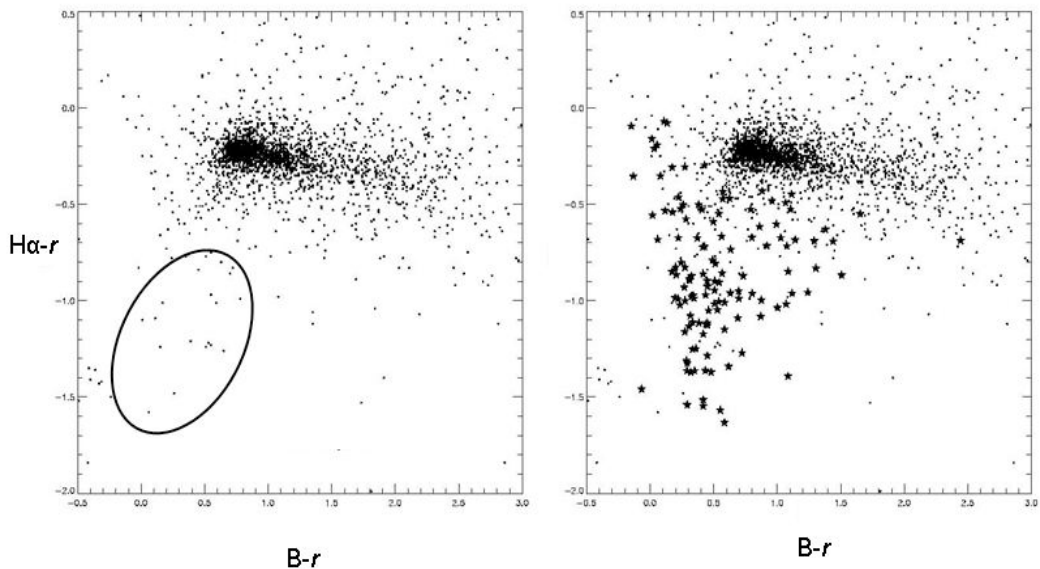

**FIGURE 6.** *Left*: multicolor *HST* photometry of 3,100 objects located within 1" of any of the *Chandra* X-ray sources in NGC 362. The oval denotes an interesting subset of blue, Hα excess stars. *Right*: identical data, but with the addition (*star symbols*) of *SDSS* photometry, converted to the *HST* system via SYNPHOT, of ~$10^2$ spectroscopically-confirmed field cataclysmic variable stars.

# 3. CONCLUSIONS

Our analysis of the NGC 362 X-ray and optical data is far from complete. While it is clear that considerably more effort will be required to establish a list of confident optical counterparts to the large number of X-ray sources in the cluster, we can already infer a number of interesting results.

a) As the cluster has of order $\sim 10^2$ X-ray sources, it is evident that the competition of formation and destruction terms for close binaries in dense clusters does not preferentially destroy large numbers of systems.
b) NGC 362 fits well on the "dense" end of the encounter frequency vs. number of source relation for globular clusters: most compact cluster binaries almost surely are dynamically formed.
c) NGC 362 has at least one qLMXB.
d) This cluster probably has a large population of cataclysmic variables and related objects, both X-ray detected and non-detected (but optically inferred).
e) Our existing data almost certainly have several dozen CVs with X-ray/optical associations, but they will be difficult to tease out individually, as they are faint and in exceptionally crowded fields.

It is clear that the seemingly vexing question of the 1980s, "where are the close binary stars in globular clusters?" was simply being asked prematurely: close binaries are present in large numbers in clusters, but are faint enough, and in such crowded fields, that the combined efforts of *Chandra* and *HST* are needed for discovery.

# ACKNOWLEDGEMENTS

Support for *Chandra* Program 05300611 was provided by NASA through the *Chandra* X-ray Observatory Center, which is operated by the Smithsonian Astrophysical Observatory for and on behalf of NASA under contract NAS8-3060. Support for *HST* Program numbers 10005 and 10615 was provided by NASA through a grant from the Space Telescope Science Institute, which is operated by the Association of Universities for Research in Astronomy, Incorporated, under NASA contract NAS5-26555. Funding for the SDSS and SDSS-II has been provided by the Alfred P. Sloan Foundation, the Participating Institutions, the National Science Foundation, the U.S. Department of Energy, the National Aeronautics and Space Administration, the Japanese Monbukagakusho, the Max Planck Society, and the Higher Education Funding Council for England. The SDSS Web Site is http://www.sdss.org/. The SDSS is managed by the Astrophysical Research Consortium for the Participating Institutions. The Participating Institutions are the American Museum of Natural History, Astrophysical Institute Potsdam, University of Basel, University of Cambridge, Case Western Reserve University, University of Chicago, Drexel University, Fermilab, the Institute for Advanced Study, the Japan Participation Group, Johns Hopkins University, the Joint Institute for Nuclear Astrophysics, the Kavli Institute for Particle Astrophysics and Cosmology, the Korean Scientist Group, the Chinese Academy of Sciences (LAMOST), Los Alamos National Laboratory, the Max-Planck-Institute for Astronomy (MPIA), the Max-Planck-Institute for Astrophysics (MPA), New Mexico State University, Ohio State University, University of Pittsburgh, University of Portsmouth, Princeton University, the United States Naval Observatory, and the University of Washington.